\documentstyle[aps,twocolumn,epsf]{revtex}

\def\half{\mbox{$1\over 2$}}

\def\beq{\begin{equation}}
\def\eeq{\end{equation}}

\def\beqa{\begin{eqnarray}}
\def\eeqa{\end{eqnarray}}
\def\lf{\nonumber \\}
\def\LP{\left(}
\def\RP{\vphantom{\half} \right)}

\newcommand{\bi}{\bibitem}
\newcommand{\Tr}{{\rm Tr \,}}

\gdef\vec#1{{\mathbf #1}}
\gdef\aver#1{\left\langle #1 \right\rangle}
\gdef\s#1{\! #1 \!}

\gdef\Eq#1{Eq.~(\ref{#1})}

\newcommand{\up}{\uparrow}
\newcommand{\down}{\downarrow}

\begin{document}
\draft
%\wideabs{
\title{Thermodynamics as an alternative foundation 
for zero-temperature \\ density functional theory 
and spin density functional theory}
\author{Nathan Argaman and Guy Makov}
\address{Physics Department, NRCN, P.O.\ Box 9001, Beer Sheva 84190, Israel}
%\date{November 5, 2001}
%\date{submitted: November 5, 2001; this version: February 17, 2002} 
\date{}
\maketitle
\begin{abstract}
Thermodynamics provides a transparent definition of the free energy of density functional theory (DFT), and of its derivatives --- the potentials, at finite temperatures $T$.  By taking the $T \to 0$ limit, it is shown here that both DFT and spin-dependent DFT (for ground states) suffer from precisely the same benign ambiguities: (a) charge and spin quantization lead to ``up to a constant'' indeterminacies in the potential and the magnetic field respectively, and (b) the potential in empty subspaces is undetermined but irrelevant.  Surprisingly, these simple facts were inaccessible within the standard formulation, leading to recent discussions of apparent difficulties within spin-DFT.
\end{abstract}
% 71.15.Mb Density functional theory, local density approximation, gradient 
% and other corrections
% 75.20.-g Diamagnetism, paramagnetism, and superparamagnetism
% 31.15.Ew Density-functional theory (atoms and molecules)
% 05.30.Fk   Fermion systems and electron gas 

Density functional theory (DFT) is the method of choice for a wide range of theoretical computations of electronic systems in chemistry, condensed matter physics, and materials science, and is also applicable to other many-body problems, such as classical fluids (for an introduction, see Ref.~\cite{AM}).  The theoretical foundations of DFT have repeatedly elicited discussion and reconsideration, e.g.\ Refs.~\cite{Levy,ensemble,vrep}, most recently in the context of spin-dependent DFT \cite{vBH,Esch,CV}.  Specifically, the first Hohenberg--Kohn theorem (HKI) states \cite{HK} that a given ground-state density distribution $n(\vec r)$ determines the corresponding external potential $v(\vec r)$ uniquely, up to an overall constant (an arbitrary reference energy).  In contrast, in spin-DFT it has been shown that the densities $n(\vec r)$ and $\vec m(\vec r)$ (the spin density) do not always determine the potentials $v(\vec r)$ and $\vec B(\vec r)$ (the magnetic field), although they do determine the ground state $\Psi$ uniquely, and hence the energy functional $F[n(\vec r),\vec m(\vec r)]$ which features in the second Hohenberg-Kohn theorem (HKII) is well-defined \cite{vBH,DG}.  Nevertheless, many developments in spin-DFT, including the main practical tool --- the spin--dependent Kohn--Sham equations \cite{KS} 
--- tacitly assume a one-to-one correspondence between densities and potentials.

DFT was immediately generalized to thermal ensembles by Mermin \cite{M}, who followed Ref.~\cite{HK} closely, proving analogues of HKI and HKII.  It was soon realized that finite-temperature DFT does not suffer from some of the ambiguities of ground-state DFT (e.g., obviously for $T>0$ a ground-state degeneracy no longer leads to a one-to-many relationship between $v$ and $n$), but only many years later it was clarified that the functional $F[n]$ (or $F[n,\vec m]$) can also be obtained at $T > 0$ by a functional extension of standard thermodynamics \cite{AM}.  Specifically, the grand potential of an electronic system $\Omega$ depends on the temperature $T$, the chemical potential $\mu$, the potential $v(\vec r)$ and the magnetic field $\vec B(\vec r)$.  A change of representation involves a Legendre transform, e.g.\ replacing $\mu$ by the electron number $N$ as a free variable.  In the case of the potentials $v(\vec r)$ and  $\vec B(\vec r)$, a functional Legendre transform must be used: $F[n]=\Omega - \int d\vec r \; v n$ for (spin-independent) DFT, and $F[n,\vec m]=\Omega - \int d\vec r \; v n + \int d\vec r \; \vec B \cdot \vec m$ for spin-DFT.

For $T > 0$, the simple one-to-one nature of the relationship between $n(\vec r)$ and $v(\vec r)$, or between the pair $n,\vec m$ and the pair $v,\vec B$, is guaranteed by the convexity argument given below.  For $T \to 0$, the ground-state formalism is regained, but the degree of convexity may also vanish in this limit, requiring special care.  The purpose of the present work is to establish this limit as an alternative foundation for zero-temperature DFT and spin-DFT.  The novel result is that HKI has essentially the same validity in ground-state spin-DFT as in DFT.

The Hamiltonian for electrons (in a large box) is
\beq
\hat H = \hat T + \Lambda \hat W + \hat V + \hat B  \; ,
\eeq
with $\hat T=(-\hbar^2 / 2 m) \sum_\sigma \int d\vec r \> 
\psi^\dagger_{\vec r \sigma} \nabla^2 \psi_{\vec r \sigma}$ 
the kinetic energy, 
$\hat W=(e^2/2) \sum_{\sigma,\sigma'} \int d\vec r \, d\vec r' \; 
\psi^\dagger_{\vec r \sigma} \psi_{\vec r' \sigma'}^\dagger 
\psi_{\vec r' \sigma'} \psi_{\vec r \sigma} / |\vec r - \vec r'|$ 
the interaction, $\Lambda=1$ a parameter introduced for later convenience, 
$\hat V=\int d\vec r \; v(\vec r) \sum_\sigma 
\psi^\dagger_{\vec r \sigma} \psi_{\vec r \sigma}$ the potential term, and 
$\hat B = -\int d\vec r \; \sum_{\sigma,\sigma'} \vec B(\vec r) \cdot
\psi^\dagger_{\vec r \sigma} \vec \tau_{\sigma\sigma'} \psi_{\vec r \sigma'}$ 
the magnetic term \cite{unc} ($\vec \tau_{\sigma\sigma'}$ is the vector of Pauli matrices, and in the units used the Bohr magneton $\mu_{\rm B} =1$).  The term $\hat B$ is optional, and distinguishes spin--DFT from DFT.   The free energy in the grand-canonical ensemble is
\beq
\Omega = -T \log \Xi  \quad ;  \quad  \Xi = \Tr \exp\LP- (\hat H-\mu \hat N)/T \RP  \; ,
\eeq
where $\Xi$ is the partition function, $\mu$ is the chemical potential, $T>0$ is the temperature (in energy units), and the total particle number operator is 
$\hat N=\sum_\sigma \int d\vec r \; \psi^\dagger_{\vec r \sigma} \psi_{\vec r \sigma}$.

The partial derivatives of $\Omega$ are the expectation value of the number of electrons, $N=-(\partial \Omega / \partial \mu)$, and the entropy $S=-\partial \Omega / \partial T$.  The functional derivatives give the density distribution 
$n(\vec r) = \delta \Omega / \delta v(\vec r)$, and for spin-DFT, the magnetic moment density 
$\vec m(\vec r) = -\delta \Omega / \delta \vec B(\vec r)$.

The grand potential $\Omega$ is a strictly concave functional of $v(\vec r)$ and $\vec B(\vec r)$.  To see this, consider the operators 
$\hat X =-(\hat H - \mu \hat N)/T$ and $\hat Y =-(\hat H' - \mu \hat N)/T$,
where $\hat H$ and $\hat H'$ incorporate different potentials, denoted $v$ and $\vec B$ for $\hat H$, and $v'$ and $\vec B'$ for $\hat H'$.  Concavity follows because for any Hermitian operators $\hat X$ and $\hat Y$ and any real $\alpha$ such that $\hat X - \hat Y \neq $const. and $0 < \alpha < 1$, one has \cite{cproof}
$\Tr \exp( \alpha \hat X + (1-\alpha) \hat Y )  < (\Tr \exp \hat X)^\alpha (\Tr \exp \hat Y)^{1-\alpha}$.

\begin{figure}[t]
%\vspace{0.5cm}
\epsfxsize=8cm
\centerline{\epsffile{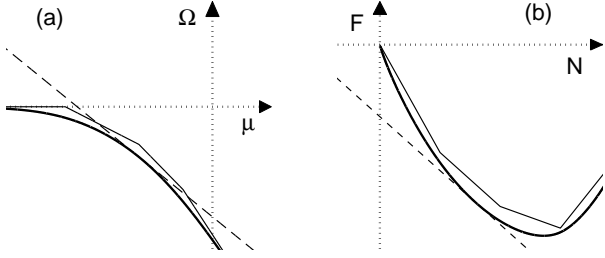}}
\vspace{0.3cm}
\caption{
(a) The grand potential $\Omega(\mu,T)$ is a concave function of the chemical potential $\mu$ (thick curve).  The Legendre transform describes this curve by its family of tangents (dashed line): their slopes $-N$ and their intercept with the vertical axis, $F=\Omega+\mu N$.  (b) The resulting $F(N,T)$ curve represents the Helmholtz free energy.  HKII corresponds to describing the intercept here as the minimal value of $F(N)-\mu N$ attainable for a given slope $\mu$, after a generalization of the variables $\mu$ and $N$ to the functional variables $\mu -v(\vec r)$ and $n(\vec r)$.  The thin lines represent the $T \to 0$ limit, for which a range of tangents can intersect at a single point (see text).
}
\end{figure}

In thermodynamics, it is often useful to switch representation by employing a Legendre transform (see Fig.~1), e.g.\ using the Helmholtz free energy $F(N,T)=\Omega(\mu,T)+\mu N$, where $\mu$ on the right hand side is chosen by the physical condition $N=-(\partial \Omega / \partial \mu)$, equivalent to maximising the expression $\Omega+\mu N$ with respect to $\mu$.
The DFT energy functional $F[n]$, generalized to spin-DFT, is similarly introduced as a functional Legendre transform
\beq \label{Leg} \!
F[n,\vec m] \! = \! \Omega[v,\vec B] \! - \!\!\! \int \!\! d \vec r \! 
\left\{ n(\vec r) \LP v(\vec r) \! - \! \mu \RP \! - \! 
\vec m(\vec r) \!\! \cdot  \!\! \vec B(\vec r) \right\} \!
, \eeq
where the right hand side is to be maximized with respect to $v$ and $\vec B$ (for DFT, $\vec m$ and $\vec B$ are simply omitted).  

The concavity of $\Omega$ guarantees both {\it existence} and {\it uniqueness} of the potentials for all reasonable density distributions, i.e.\ smooth distributions with $n(\vec r) > |\vec m(\vec r)|$ for spin-DFT, and $n(\vec r) > 0$ for DFT.  The conditions of $v$-representability or $N$-representability, which are necessary in the conventional approach \cite{vrep,HK}, do not arise here.  {\it Uniqueness}: as the right hand side of \Eq{Leg} is concave in $v$ and $\vec B$, it cannot have more than one maximum.  This replaces the standard {\it reductio ad absurdum} argument of DFT, based on the Rayleigh-Ritz minimum energy principle (or a generalization thereof), and used to prove HKI \cite{HK,M}.  {\it Existence}: the maximization of \Eq{Leg} can fail only if arbitrarily large values are obtainable on the right hand side (e.g.\ for unreasonable density distributions with $n<0$, very large values of $v-\mu$ can be taken with $\Omega$ negligible).  Any reasonable $n,\vec m$ set can be obtained as a weighted sum of distributions for which solutions are known to exist.  As the right hand side of \Eq{Leg} is linear in the densities, its value can never be larger than the weighted sum of the values of $F$ corresponding to these known distributions --- $F[n,\vec m]$ is convex.  The relationship between potentials and densities at $T > 0$ is thus one-to-one.  It follows that the maximum over $v$ and $\vec B$ in \Eq{Leg} is obtained for the physical system for which $\delta \Omega / \delta v(\vec r) = n(\vec r)$ and 
$\delta \Omega / \delta \vec B(\vec r) = \vec m(\vec r)$.

The free-energy functional $F$, in the combination 
\beq \label{iLeg}
F[n,\vec m] + \int d \vec r \, \left\{ n(\vec r) \LP v(\vec r)-\mu \RP -
 \vec m(\vec r) \cdot \vec B(\vec r) \right\}
\;  ,  \eeq
can be minimized with respect to the density distributions to give back the grand potential $\Omega$.  This is the inverse Legendre transform in thermodynamics, and corresponds to HKII in DFT \cite{HK}.  The derivatives here are $\delta F/\delta n = \mu - v$ and 
$\delta F/\delta \vec m = B$, as follows from \Eq{Leg}.  The other derivatives of $F[n,\vec m]$ also follow, e.g.\ the entropy $\partial F/\partial T = \partial \Omega/\partial T = -S$, where as usual the derivatives of $F$ are taken at constant $n(\vec r)$ and $\vec m(\vec r)$, while those of $\Omega$ are taken at constant $\mu$, $v(\vec r)$ and $\vec B(\vec r)$ (here $F[n,\vec m]$ does not depend on $\mu$ because the chemical potential appears in $\Omega$ only in the combination $v-\mu$).

The derivative $\partial F/\partial \Lambda = 
\partial \Omega / \partial \Lambda = \langle \hat W \rangle$ 
is of special interest, as $F[n,\vec m] = 
F_{\rm ni}[n,\vec m] + \int_0^1 d\Lambda \; (\partial F/\partial \Lambda)$, 
makes the connection \cite{adiabatic} with the Kohn--Sham noninteracting system, described by $F_{\rm ni}$ which is $F$ for $\Lambda=0$.  The expectation value of the interaction operator, $\langle \hat W \rangle$, is evaluated here for that system which has the density distributions $n$ and $\vec m$ and a reduced interaction strength $\Lambda$.  The major part of the interaction energy is given directly by the density $n$ as the Hartree term 
$E_{\rm H}[n]=\int d\vec r \, d\vec r' \; e^2 \, n(\vec r) n(\vec r') / 2|\vec r - \vec r'|$.
Thus 
\beq \label{nic}
F[n,\vec m] = F_{\rm ni}[n,\vec m] + E_{\rm H}[n] + F_{\rm xc}[n,\vec m]
\eeq
where $F_{\rm xc}$ is the exchange-correlation (xc) energy,
\beq \label{Fxc}
F_{\rm xc}[n,\vec m] = \int d\vec r \> f_{\rm xc}[n,\vec m](r)  \; ,
\eeq
with $f_{\rm xc}(r)$, the xc energy density, defined as 
\beq \label{grr}
f_{\rm xc}[n,\vec m](r) = \int_0^1 d\Lambda \sum_{\sigma,\sigma'}  
\int { e^2 \, n_\sigma(\vec r) d\vec r' \over 2|\vec r - \vec r'|}
 \rho^{\rm xc}_{\sigma,\sigma'}(\vec r, \vec r') \; ,
\eeq
in terms of the so-called xc-hole density 
\beq \label{rhoxc}
\rho^{\rm xc}_{\sigma,\sigma'}(\vec r, \vec r') = 
{1 \over n_\sigma(\vec r)} 
\langle \psi^\dagger_{\vec r \sigma} \psi_{\vec r' \sigma'}^\dagger 
\psi_{\vec r' \sigma'} \psi_{\vec r \sigma} \rangle - n_{\sigma'}(\vec r')  \; ,
\eeq
where $n_{\sigma}(\vec r) = \half\LP n(\vec r) + \sigma m_z(\vec r) \RP$, $\sigma=\pm 1$ (the sum-rule 
$\int d\vec r' \; \rho^{\rm xc}_{\sigma,\sigma'}(\vec r, \vec r') = 
- \delta_{\sigma,\sigma'}$ holds at $T=0$).  The 
${\delta \over \delta n(\vec r)}$ and 
${\delta \over \delta \vec m(\vec r)}$ derivatives of \Eq{nic} give
\beqa \label{KSpot}
\mu - v(\vec r) & = & \mu - v_{\rm ni}(\vec r) -e\varphi(\vec r)+v_{\rm xc}(\vec r)  \lf
\vec B(\vec r) & = & \vec B_{\rm ni}(\vec r) - \vec B_{\rm xc}(\vec r)
\eeqa
in obvious notation, retrieving the results of Kohn and Sham \cite{KS} and generalising them to spin-DFT.  In practice, one requires approximations of $F_{\rm xc}$ or $f_{\rm xc}$, which allow their derivatives to be taken, e.g.\ the local spin-density approximation where 
$f_{\rm xc}[n,\vec m](r)$ depends only on $n(\vec r)$ and $\vec m(\vec r)$.  With explicit expressions for $v_{\rm xc}$ and $\vec B_{\rm xc}$ thus obtained (and noninteracting physics used to relate $n$,$\vec m$ with $v_{\rm ni}$,$\vec B_{\rm ni}$), \Eq{KSpot} yields a very powerful and simple scheme (simpler than Hartree), for computations of $n$, $\vec m$ and $\Omega$ of interacting electrons in given external fields.  Eqs.~(\ref{Fxc}---\ref{grr})
for $F_{\rm xc}$ in terms of $\rho^{\rm xc}$ form the basis for most discussions and improvements of the accuracy of DFT,
as e.g.\ in the recent Ref.~\cite{Kohn_ens}, which suggests a modified LDA for cases involving degeneracy.

At zero temperature, $\Omega$ and $F$ are defined by their (well-behaved) 
$T \to 0$ limits: $\Omega$ tends to the minimal eigenvalue of $\hat H - \mu \hat N$, and its Legendre transform gives $F$, which at $T=0$ is the internal energy.  Two issues require consideration: (a) the degree of concavity or convexity vanishes as $T \to 0$, resulting in discontinuities (indeterminancy) in the derivatives; and (b) interest in fully polarized systems, $|\vec m(\vec r)| = n(\vec r)$, becomes legitimate.

Consider first fully polarized systems, defining the value of $F[n,\vec m]$ at $|\vec m(\vec r)|=n(\vec r)$ as the limit of its $T=0$ values as $|\vec m(\vec r)| \to n(\vec r)$.  When $\vec B(\vec r)$ and $\vec m(\vec r)$ are everywhere parallel to $z$, as is very often assumed in spin-DFT, the spin-dependent densities $n_{\up\down}=\half(n \pm m_z)$ and potentials $v_{\up\down}= \half(v \mp B_z)$ are useful; aligning all spins in (say) the up direction leaves $v_\down$ undetermined (for $T>0$, this potential diverges as $n_\up \to n$).  Indeed, the whole $v_\down(\vec r)$ {\it function} may vary without changing the state of the system, provided only that the lowest-lying spin-down state remains above the energies of the spin-up electrons present \cite{CV}.  However, having $v_\down(\vec r)$ undetermined poses no difficulty, because that subspace is empty, $n_\down(\vec r)=0$.  The analogue of this in spin-independent DFT is the situation with $n(\vec r) \equiv 0$ for $\vec r$ in a finite region ($N=0$ if this region covers the whole system) --- the potential within this region is undetermined, and one may exclude it from the system.  It is of little significance that in spin-DFT the density in a subspace may vanish without any divergence in the potentials (at $T=0$), whereas in DFT a region with $n(\vec r) \equiv 0$ may require $v \to \infty$ on (or near) its boundary.  As shown in Ref.~\cite{vBH}, in the one-electron case this situation persists for non-parallel magnetic fields.

Consider next a thermodynamic example of vanishing convexity (see Fig.~1): at $T=0$, the relationship between $N$ and $\mu$ is step-wise \cite{PPLB} --- a range of $\mu$ from the ionization potential to the electron affinity corresponds to a single value of $N$, while all $N$ between $N_0$ and $N_0+1$ share a single value of $\mu$.  Charge is quantized, due to particle-number conservation, $[\hat N , \hat H]=0$.  The system switches between ``rigid'' states when $\mu$ is varied, with a ``soft mode'' occuring at each switching event --- the relevant susceptibility alternates between infinite and vanishing values.  Rigidity might affect applications of DFT, e.g.\ indeterminancy of $\mu$ for integer $N$ in \Eq{KSpot}.

In its original form, DFT deals with electron number quantization by restricting attention to a subspace of fixed $N$.  The functional derivative $\delta F/\delta n$ is redefined in terms of density variations with $\int d\vec r \; \delta n(\vec r) = 0$, and contains an overall arbitrary constant, corresponding to a choice of reference energy for the potential $v(\vec r)$, or to adding a multiple of $\hat N$ to $\hat H$.  Alternatively, one may fix the chemical potential $\mu$, and define the derivatives through {\it their} $T \to 0$ limit (barring degeneracies, this is the midpoint of the discontinuity).  The derivative $\delta F/\delta n$ at integer $\int d\vec r \; n = N$ then gives an allowable choice of $v(\vec r)-\mu$, but choices differing from it by a (small enough) constant give the same $n(\vec r)$.  Similarly, $\delta \Omega/\delta v$ evaluated for a degenerate $v$ gives the ensemble averaged $n$, but any weighted combination of the densities of the individual ground states also corresponds to $v$.

For parallel-spin DFT, the $z$ component of the total spin is a second conserved, quantized quantity, $[\hat M_z,\hat H]=0$.  Again, one can limit attention to a subspace with a given value of $M_z$ (and of $N$), requiring $\int d\vec r \; \delta m_z = 0$.  Shifting the magnetic field $B_z(\vec r)$ by a constant --- making an ``arbitrary choice of the origin of reference magnetic field'' --- corresponds to adding a multiple of $\hat M_z$ to $\hat H$ and cannot alter the state of the system as long as $M_z$ is fixed.  The alternative of allowing $M_z$ to vary reveals the succession of discontinuous switching events between rigid spin states.  Clearly, the $n(\vec r),m_z(\vec r)$ distributions of each of these states correspond to a range of potentials, obtainable by shifting $v(\vec r)-\mu$ and $B_z(\vec r)$  by {\it two} distinct constants (small enough not to change $N$ or $M_z$).

One may argue on physical grounds that a linear segment in $\Omega$ --- a rigid state --- will not occur (for $|\vec m|<n$) without adjacent cusps.  Such a linear segment in $\Omega$ corresponds to a cusp or derivative-discontinuity in $F$.  Each adjacent cusp in $\Omega$, representing a ground-state degeneracy or soft mode, is accompanied by quantization of some physical quantity $\hat O$ which commutes with $\hat H$.  In the cases considered above, the operator $\hat N$ or $\hat M_z$ can be constructed from the potential terms.  When the operator $\hat O$ cannot be so constructed, e.g.\ the angular momentum operator $\hat L$ for a spherically symmetric system, a soft mode or cusp in $\Omega$ may result, but rigidity (a cusp in $F$) can not, precisely because no combination of potentials is able to ``push'' the system in its rigid direction.

The arguments used here are general, and apply to other extensions of DFT.  For example, it is possible to construct $\hat L$ from the potentials of current-DFT; for density distributions with the quantized values of electron number, spin and angular momentum, the potentials are then determined up to {\it three} terms \cite{CVnew}, corresponding to adding multiples of each of these operators to $\hat H$.  A simpler example is a spin-singlet state with $\vec m(\vec r) \equiv 0$ in spin-DFT, where the corresponding Hamiltonian may conserve all three components of $\hat \vec M$.  The magnetic field is then undetermined in both magnitude and direction: a small constant-$\vec B$ term will not change $n(\vec r)$ and $\vec m(\vec r)$.

Additional support for our conclusions is provided by the fact that generating atypical exceptions requires tuning many more parameters than are available.  Indeed, all existing counter-examples \cite{vBH,Esch,CV} to the spin extension of HKI are either quantized or fully-polarized states.  In an apparent exception, Ref.~\cite{Esch} identifies conditions which allow a magnetic field $\Delta \vec B(\vec r)$ of constant magnitude but space-dependent direction to be added to $\hat H$ without changing $n$ and $\vec m$ (and $\psi$), even for $|\vec m|<n$.  However, these conditions cannot be fulfilled by a many-electron ground state \cite{caseB}.

The thermodynamic point of view sheds new light on each of the five issues raised in Ref.~\cite{CV}:  (i) {\it Excited states} are accessible to DFT and spin-DFT as thermal ensembles, for which the one-to-one nature of the potentials---densities relationship is guaranteed.  (ii) Construction of {\it accurate xc potentials} from Monte Carlo or configuration interaction calculations can proceed in spin-DFT as in conventional DFT, provided one acknowledges the ``up to a constant'' nature of both the external potential and the magnetic field, and the irrelevance of potentials in empty subspaces.  (iii) The presence of an {\it excitation--gap} need not lead to indeterminate potentials.  (iv) However, {\it band gaps} in semiconductors and in half-metallic ferromagnets cause a zero--temperature first--order phase transition: the (spin dependent) chemical potential is discontinuous as a function of filling.  Such derivative discontinuities are of different magnitudes (and can occur at different density distributions) for the interacting and noninteracting cases, and devising approximations to $F_{\rm xc}$ which account for them remains a challenge.  Progress may be achieved by identifying the bands to which the electron densities in Eqs.~(\ref{Fxc}---\ref{rhoxc}) belong.  Obviously, other electronic phase transitions will pose difficulties as well, at least in the thermodynamic limit.  (v) Finally, {\it functional derivatives} can be taken either at $T \to 0+$, or at fixed $N$ and $\vec M$, with $\int d\vec r \> \delta n = \int d\vec r \> \delta \vec m = 0$.  Chain rules, as in the optimized-effective potential method, then hold.

To summarize, thermodynamic considerations followed by the $T \to 0$ limit can serve as an alternative foundation for DFT and its spin-dependent extension.  At finite temperatures, the relationship between the potentials and the densities is guaranteed to be one--to--one by the strict concavity of the grand potential $\Omega$, but for $T \to 0$ linear segments may appear in $\Omega$, due to quantization of spin and charge.  Consenquently, in the $T=0$ ground-state theory the densities determine the potentials (in non-empty subspaces) up to {\it two} spatial constants, one in the external potential and one in the magnetic field.  The fact that more complicated ambiguities do not arise, often assumed by practitioners, has been shown here for the first time.  One may either work at finite temperatures or fix the total electron number and the total spin and work within a subspace.  Either option resolves all the difficulties of principle raised in Refs.~\cite{Esch,CV}; the remaining difficulties are practical, involving the construction of accurate xc approximations in cases of complex physical behavior (electronic phase transitions).

The application of thermodynamic considerations to DFT may serve as an example of the unity of physics.  It avoids some of the pitfalls encountered in the analysis of spin-DFT along the lines of the original Hohenberg--Kohn theorems.  It would be interesting to compare these developments also to a detailed analysis of spin-DFT within the constrained-search approach \cite{Levy,ensemble}.

\end{document}